\documentclass[prl,aps,twocolumn,superscriptaddress,showpacs]{revtex4-1}

\usepackage{graphicx}
\usepackage{epstopdf}

\begin{document}

\title{Soft striped magnetic fluctuations competing with superconductivity in Fe$_{1+x}$Te}

\author{C. Stock}
\affiliation{School of Physics and Astronomy, University of Edinburgh, Edinburgh EH9 3JZ, UK}
\author{E. E. Rodriguez}
\affiliation{Department of Chemistry of Biochemistry, University of Maryland, College Park, MD, 20742, U.S.A.}
\author{O. Sobolev}
\affiliation{Forschungs-Neutronenquelle Heinz Maier-Leibnitz, FRM2 Garching, 85747, Germany}
\author{J. A. Rodriguez-Rivera}
\affiliation{NIST Center for Neutron Research, National Institute of Standards and Technology, 100 Bureau Dr., Gaithersburg, MD 20889}
\affiliation{Department of Materials Science, University of Maryland, College Park, MD  20742}
\author{R.A. Ewings}
\affiliation{ISIS Facility, Rutherford Appleton Laboratory, Didcot, OX11 0QX, U.K.}
\author{J.W. Taylor}
\affiliation{ISIS Facility, Rutherford Appleton Laboratory, Didcot, OX11 0QX, U.K.}
\author{A. D. Christianson}
\affiliation{Quantum Condensed Matter Division, Oak Ridge National Laboratory, Oak Ridge, Tennessee 37831, USA}
\author{M. A. Green}
\affiliation{School of Physical Sciences, University of Kent, Canterbury, CT2 7NH, UK}

\date{\today}

\begin{abstract}

Neutron spectroscopy is used to investigate the magnetic fluctuations in Fe$_{1+x}$Te - a parent compound of chalcogenide superconductors.  Incommensurate ``stripe-like" excitations soften with increased interstitial iron concentration.  The energy crossover from incommensurate to stripy fluctuations defines an apparent hour-glass dispersion.  Application of sum rules of neutron scattering find that the integrated intensity is inconsistent with an $S$=1 Fe$^{2+}$ ground state and significantly less than $S$=2 predicted from weak crystal field arguments pointing towards the Fe$^{2+}$ being in a superposition of orbital states.  The results suggest that a highly anisotropic order competes with superconductivity in chalcogenide systems.  


\end{abstract}

\pacs{}

\maketitle

Pnictide and chalcogenide superconductors have altered the view of what provides the basis for high temperature superconductivity.   While the cuprate superconductors universally derive from Mott insulators which can, at least qualitatively, be understood in terms of a single electronic band, the parent phase of iron based superconductors has been less clear: Fe-based parent phases are either poorly metallic or semimetallic resulting in a debate over whether a localized or itinerant/spin density wave picture is more appropriate.~\cite{Johnston10:59,Birgeneau06:75}  Towards this goal, it is important to understand the magnetic excitation spectrum in starting materials as superconducting variants consist of fluctuating versions of this ground state.~\cite{Lee06:78}  Here we study Fe$_{1+x}$Te which is arguably the structurally simplest of the iron superconductors based upon single layers of tetrahedrally coordinated Fe$^{2+}$ ions.~\cite{Fruchart75:10,Hsu08:105}    While the iron superconductors have been shown to display both localized~\cite{Si08:101,Yao08:78} and itinerant properties~\cite{Mazin08:101,Kuroki08:101}, Fe$_{1+x}$Te hosts one of the most localized responses of all iron based superconductors evidenced by large ordered magnetic moments and calculated heavy band masses.~\cite{Yin11:10}  

In this study,  we combine neutron scattering data from spectrometers with overlapping dynamic ranges on two samples of Fe$_{1+x}$Te to understand the magnetic fluctuations.  We report a one dimensional incommensurate excitation that softens with increased charge doping with interstitial iron and hence competes with unconventional chalcogenide superconductivity.  We apply sum rules of neutron scattering to evaluate the spin and orbital ground state of the iron cations.  The results represent a dynamical signature of a highly anisotropic striped order which competes with superconductivity in the chalcogenides.

\begin{figure}[t]
\includegraphics[width=8.75cm] {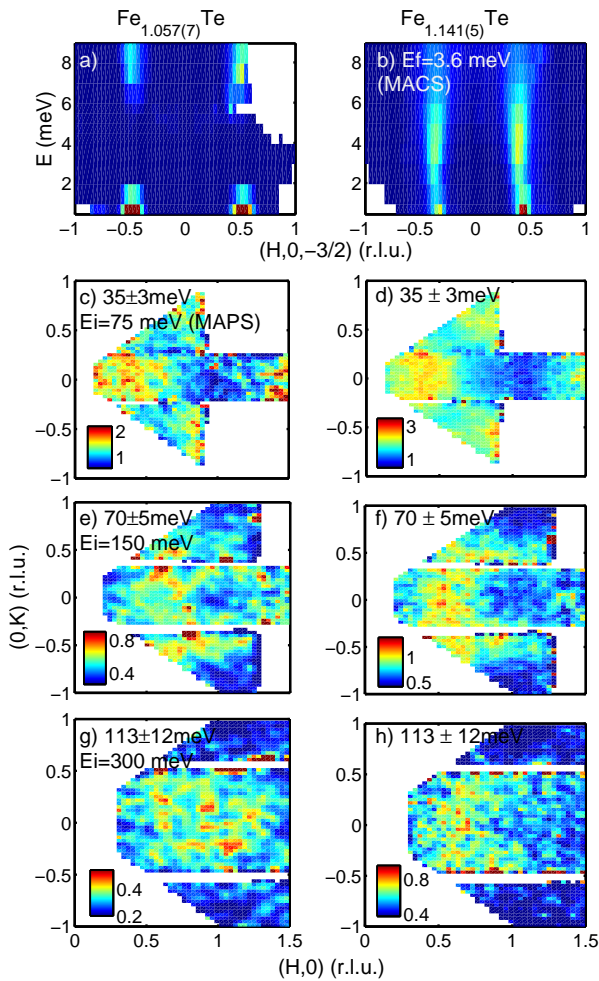}
\caption{\label{summary} Neutron inelastic scattering comparing commensurate Fe$_{1.057(7)}$Te with incommensurate Fe$_{1.141(5)}$Te.  $a-b)$ Constant-Q slices of the low-energy data taken on MACS (E$_{f}$=2.6 meV).  $c-h)$ Constant energy slices from MAPS taken with incident energies of 75, 150, and 350 meV. }
\end{figure}

Superconductivity in Fe$_{1+x}$Te$_{1-y}$Ch$_{y}$ (where Ch is a chalcogenide ion) has been most commonly achieved through anion substitution on the Te site $y$ with either sulfur or selenium.~\cite{Sales09:79,Mizuguchi09:94}  However, the cation concentration (interstitial iron $x$) in Fe$_{1+x}$Te$_{1-y}$Ch$_{y}$ is directly correlated with the anion concentration ($y$) and chemical techniques have been developed to independently tune $x$ and $y$.~\cite{Rodriguez10:132}  Several studies have found that changing the concentration of interstitial iron has analogous effects to anion doping for a fixed selenium concentration.~\cite{Tsyrulin12:14,Rodriguez11:2,Stock12:85}  

The structural and magnetic properties of Fe$_{1+x}$Te as a function of $x$ have been reported by several groups giving generally consistent results.~\cite{Rodriguez11:84,Koz13:88,Rossler11:84,Mizuguchi12:152,Zaliznyak12:85}  A neutron diffraction study found a phase diagram with two distinct phases as a function of interstitial iron.~\cite{Rodriguez13:88}  For low concentrations of $x<$11\%, a commensurate collinear magnetic phase is realized with the critical properties being first order.  For large  $x>$11\%, the transition is second order with a spiral magnetic low temperature phase.  The two extremes are separated by a tricritical-like point at $x\sim$11\% where short-range incommensurate spin-density wave order is observed.  Resistivity measurements found the collinear $x<$11\% values to be metallic at low temperatures while larger $x>$11\% are ``semi" (or poorly) metallic and scattering from incommensurate spin fluctuations was implicated as the origin of the enhanced resistivity.~\cite{Chen09:79}  Therefore, based upon the fixed selenium studies~\cite{Rodriguez11:2,Stock12:85} and these magnetic and structural results, metallicity and superconductivity are favored for smaller values of interstitial iron.   

Doping charge through interstitial iron therefore remains an independent means of controlling the electrical properties of the chalcogenides.  We present neutron inelastic data taken from steady state reactor sources (MACS, PUMA, and HB1) and time of flight instruments (MAPS) based at pulsed spallation sources.  Further experimental and sample details are given in the supplementary information (see also Ref. \onlinecite{Rodriguez08:19}) and also details on phonon contamination and how these were disentangled (see also Ref. \onlinecite{Fong96:54}).

\begin{figure}[t]
\includegraphics[width=8.5cm] {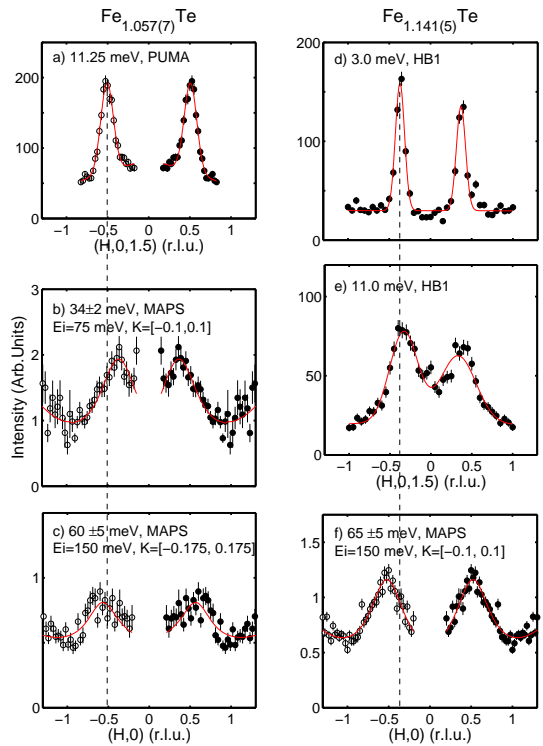}
\caption{\label{cuts} Cuts along the (H, 0) direction for commensurate $x$=0.057(7) ($a-c$) and incommensurate $x$=0.141(5) ($d-f$) crystals.  The data is taken from thermal triple-axis and spallation source data.  Open circles are symmetrized data and displayed for visual comparison.}
\end{figure}

We first describe the dispersion of the spin excitations in momentum.  Representative constant energy and momentum slices are displayed in Fig. \ref{summary} for both interstitial iron concentrations.  Panels $a-b)$ show high energy resolution constant-$Q$ slices  illustrating the gapped nature of the excitations for collinearly ordered $x$=0.057(7) and the gapless low-energy incommensurate fluctuations for $x$=0.141(5) in the spiral magnetic phase.~\cite{Stock11:84}  Higher energy excitations are displayed in panels $c-h)$ through a series of constant energy slices at 35, 70 and 113 meV.  The data do not show clean circular spin-wave cones, but rather excitations broad in momentum and dispersing to the zone boundary.

\begin{figure}[t]
\includegraphics[width=8.7cm] {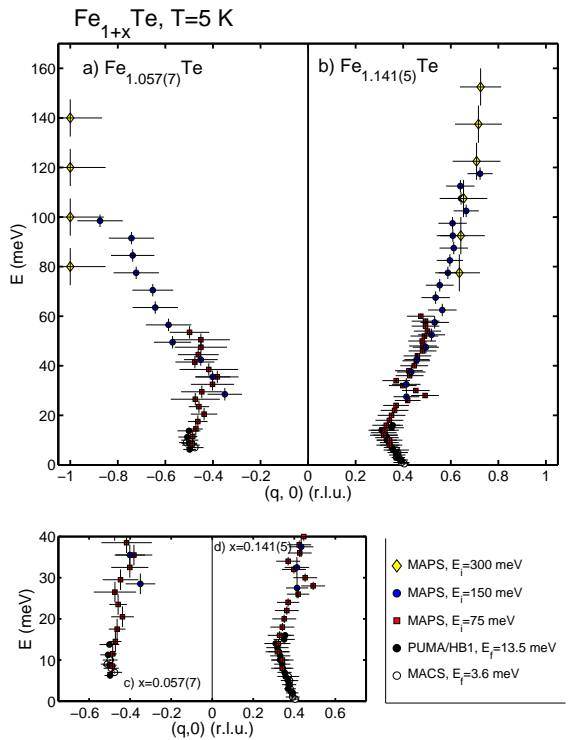}
\caption{\label{dispersion} The dispersion of the magnetic excitations in Fe$_{1+x}$Te with $x=0.057(7)$ ($a$) and $x=0.141(5)$ representative of commensurate and incommensurate magnetically ordered phases respectively.  $c)$ and $d)$ show the low-energy portion of the magnetic dispersion.  The data is a compilation of triple-axis and spallation data.  The inward dispersion described in the text is highlighted by the arrows.} 
\end{figure}

Constant energy cuts for both commensurate and incommensurate crystals are presented in Fig. \ref{cuts}.   The solid lines are fits to a gaussian lineshape multiplied by the isotropic Fe$^{2+}$ form factor squared~\cite{Brown:tables} from which a peak position and integrated intensity were derived.    The open symbols are symmetrized data around $\vec{Q}$=0.  The vertical dashed lines emphasize the fact that as the energy transfer is increased, the peak position in momentum disperses inward and then outward at higher energies.

Based upon fits (Fig. \ref{cuts}) we construct a dispersion curve (Fig. \ref{dispersion}) along the $[1,0]$ direction comparing the commensurate and incommensurate crystals.  For the commensurate $x$=0.057(7) sample, the excitations are gapped (Fig. \ref{summary}) and then disperse inwards to a wave vector $H \le$0.4 at around 30 meV energy transfer.    The excitations then disperse towards the zone boundary which is reached at $\sim$ 100 meV. Interestingly, the excitations are nearly vertical as they extend to higher energies indicative of strong dampening at the zone boundary.   This marks a clear distinction from predictions based upon a localized Heisenberg exchange.~\cite{Lipscombe11:106}  A strong zone boundary dampening has been reported in superconducting iron based variants~\cite{Lumsden10:6}, in cuprates and associated with the onset of the electronic pseudogap~\cite{Stock10:82}, and predicted to exist in Cr metal.~\cite{Sugimoto13:87}  Our results however mark a clear difference between parent cuprates (and even cuprates close to the charged doped boundary of superconductivity)~\cite{Coldea86:01,Stock07:75} and iron based systems as we do not see localized spin-waves which can be interpreted in terms of a localized Heisenberg model on a Mott insulating ground state.

The fluctuations in incommensurate $x$=0.141(5) are different.  The low-energy fluctuations are gapless and disperse inwards until $\sim$ 20 meV and then disperse outwards until the highest energy transfers studied.  However, in contrast to the commensurate $x$=0.057(7) material the excitations do not reach the zone boundary but disperse up to the highest energies studied.  There are therefore two effects of doping with interstitial iron - first, to decrease or soften the inward dispersing minimum in the magnetic excitations, and second, to increase the top of the excitation band.  A common feature from both interstitial iron concentrations is the inward (or nearly vertical) dispersion at low energies.  This dispersion never reaches the commensurate $Q$=0 positions, but disperses towards an incommensurate position that softens in energy with increased interstitial iron concentration.   

\begin{figure}[t]
\includegraphics[width=9.2cm] {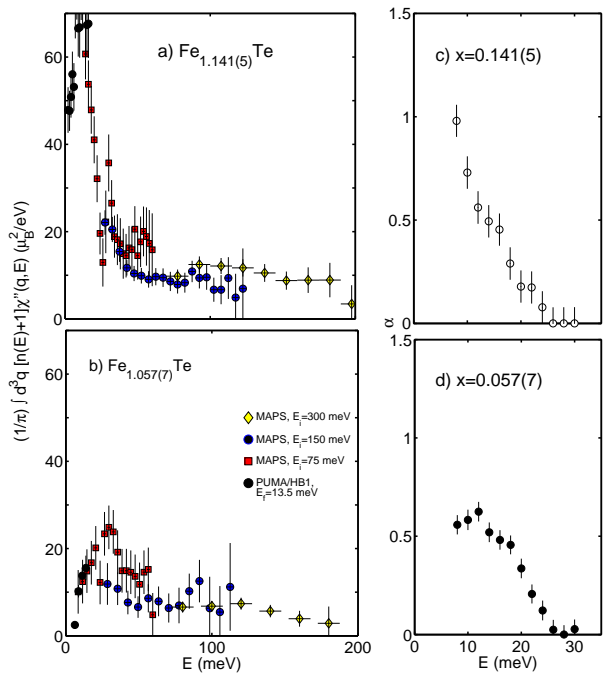}
\caption{\label{integrated} Momentum integrated intensity in absolute units as a function of energy transfer for $a)$ incommensurate $x$=0.141(5) and $b)$ commensurate $x$=0.057(7) samples.  The ``stripe" correlation parameter $\alpha$ is plotted for both concentrations in $c-d)$.}
\end{figure}

The energy inward dispersion in Fig. \ref{dispersion} also represents a cross over from two-dimensional excitations to strongly one-dimensional where the momentum dependence is well defined in H, however broad along both L and K.  This is illustrated in Fig. \ref{summary} $f$ that show the magnetic fluctuations form stripes at high energy.  To characterize this, we have fit the K dependence at each energy transfer to $I(K)\propto (1+2\alpha \cos(\vec{Q}\cdot \vec{b}))$, where $\alpha$ represents the strength of correlations between stripes aligned along $a$.  The results are shown in Fig. \ref{integrated} $c-d)$ for both interstitial iron concentrations.  The parameter $\alpha$ falls, within error, to zero at energies above the inward dispersion indicating one dimensional stripy fluctuations - a feature absent in superconducting chalcogenides~\cite{Lumsden10:6} and pnictides~\cite{Diallo09:102,Zhao09:5,Lester10:81}.   The highly one-dimensional nature of the fluctuations indicate that a highly anisotropic order is proximate in the chalcogenide superconductors.  The anisotropy exceeds that observed in superconducting LaFeAsO~\cite{Rama13:87} and CaFe$_{2}$As$_{2}$ in the paramagnetic phase~\cite{Diallo81:10}.  While the the results may indicate stripe-like fluctuations, as discussed in the context of the cuprates,~\cite{Uhrig04:93,Carlson04:70}  it may also reflect an underlying anisotropy associated with the orbital ground state.  It is difficult to interpret the results in terms of anisotropic localized exchange (as recently done for the low energy fluctuations K$_{0.85}$Fe$_{1.54}$Se$_{2}$~\cite{Zhao14:112} and SrCo$_{2}$As$_{2}$~\cite{Jayasekara13:111}) given the lack of spin-wave cones and the integrated intensity discussed below.  Highly anisotropic orders such as quadruopolar order, discussed in terms of triangular $S$=1 magnets originating from biquadratic exchange (term ``spin nematic"),~\cite{Tsunetsugu06:75,Lauchli06:97} or ``nematic" order connected with the underlying Fermi surface topology may be the origin.~\cite{Ivanov03:68,Fernandes14:10,Tsuchiizu13:111,Ohno13:82}  We note that all of these proposals predict a director where, in analogy to liquid crystals, there is some form of orientational order.   

To understand the underlying ground state, we now discuss the integrated intensites.  The $\vec{q}$ integrated intensity for the commensurate and incommensurate samples are shown in Fig. \ref{integrated}.  The calibration method is discussed in the supplementary information (see also Ref. \onlinecite{Stock04:69,Shirane:book,Fong00:61,Dai13:88,Xu13:84}).  The integrated intensity shows a peak near where the momentum dispersion (Fig. \ref{dispersion}) shows a minimum in wavevector reflecting a van-Hove type singularity where the group velocity of the magnetic mode reaches zero.  For both interstitial concentrations at large energy transfers above the peak in the local susceptibility, the integrated intensity is nearly constant.  The average value at these energy transfers is similar to the normalized values reported for pnictide systems such as CaFe$_{2}$As$_{2}$~\cite{Zhao09:5} indicating a strong similarity in the physics between the pnictides and the chalcogenides.  

The combined inward dispersion and peak in the local susceptibility indicate an hour-glass like dispersion.  While similar to La$_{2-x}$Sr$_{x}$CuO$_{4}$, the momentum dependence in Fe$_{1+x}$Te differs to YBa$_{2}$Cu$_{3}$O$_{6+x}$ where the two branches meet at the commensurate ($\pi$, $\pi$) point.~\cite{Christensen04:93,Tranquada04:429,Reznik08:78,Stock05:71}  Similar structures have also been observed in more localized La$_{5/3}$Sr$_{1/3}$CoO$_{4}$~\cite{Boothroyd11:472} and single-layered manganites~\cite{Ulbrich12:108}.  Interestingly, the hour-glass dispersion is absent in the superconducting state Fe$_{1+x}$Te$_{0.7}$Se$_{0.3}$.~\cite{Chi11:84}   An analogous ``U" type dispersion was observed to be stabilized by Ni/Cu doping~\cite{Xu12:109} which suppressed superconductivity and incommensurate order has been observed near the superconducting phase in BaFe$_{2-x}$Ni$_{x}$As$_{2}$ and Fe$_{1+x}$Se$_{y}$Te$_{1-y}$.~\cite{Luo12:108,Khasanov09:80}  These results indicate that the soft incommensurate mode is detrimental to superconductivity and, given the presence in both localized and metallic magnets, that the hour-glass dispersion is not directly tied to an electronic origin.  Our results would point towards the hour-glass point marking a cross over from two dimensional to one dimensional fluctuations as discussed above.

Based on the available magnetic and crystal field data, it is not clear how to understand the single-ion properties of the tetrahedrally coordinated Fe$^{2+}$ ion in Fe$_{1+x}$Te. Hence the neutron scattering cross section, which is typically fixed by the value of $S$, is uncertain.  In a localized model, there are two possible scenarios for populating the 3d$^{6}$ electron configuration.~\cite{Si08:101,Cao08:77}  In the the first case, termed the weak or intermediate crystal field limit, the Hund's energy scale dominates and the low-energy doubly degenerate $| e \rangle$ and higher energy triply degenerate $| t\rangle$ states are populated giving $S$=2.  The other extreme, referred to as the large crystal field limit, the energy splitting between $| e \rangle$ and $| t \rangle$ dominates and this results in an orbital triplet state with $S$=1.~\cite{Kruger09:79}   An interplay between these two energy scales has been suggested to cause a possible spin-state transition in pnictides.~\cite{Gretarsoon13:110}  Fe$_{1+x}$Te has been argued to be in this strong crystal field $S$=1 limit.~\cite{Turner09:80,Haule09:11}  This also seems to corroborated by a series of neutron diffraction results in the chalcogenide and pnictide systems where small ordered (proportional $gS$, where $g$ is the Lande factor) moments are reported.

The neutron scattering cross section is governed by several sum rules and in particular the zeroeth moment sum rule which can be written as  $\int dE \int d^{3}q {1\over \pi} [n(E)+1] \chi''(q,E)={1\over 3}g^{2}\mu_{B}^{2} S(S+1)$ (further details provided in the supplementary information).  The integral includes both elastic and inelastic scattering contributions and is independent of broadening due to itinerant effects as the integral is performed over all momentum and energy.  Some estimates on the value for $S$ have been made based upon purely localized spin-wave models as in CaFe$_{2}$As$_{2}$~\cite{Zhao09:5} and BaFe$_{2}$As$_{2}$~\cite{Ewings08:78}. These  have been summarized for other 122 systems and are typically in the range from  $S\sim$0.4-1.~\cite{Schmidt10:81}  Pure FeAs has an ordered spiral magnetic moment of only 0.5 $\mu_{B}$ with no dynamics reported.~\cite{Rodriguez11:83}   These small values are consistent with a strong crystal field picture fixing $S$=1.   However, we also note that neutron inelastic scattering results on Mott insulating La$_{2}$O$_{2}$Fe$_{2}$OSe$_{2}$ have been consistent with the weak crystal field picture with $S$=2~\cite{Free10:81,Zhu10:104,McCabe:xx} and the large ordered moments in K$_{x}$Fe$_{2-y}$Se$_{2}$ variants.~\cite{Zhao12:109} 

Through the use of the total moment sum rule we can estimate $S$ in Fe$_{1+x}$Te.  As we have noted, while our results which extend up to energy transfers of 175 meV do not capture all of the magnetic cross section, an integral over this energy range gives a lower limit on the total spectral weight and hence an effective $S$.    Combining both static and dynamic contributions gives 3.4 $\pm$ 0.3 $\mu_{B}^{2}$ and 3.7 $\pm$ 0.3 $\mu_{B}^{2}$ for $x$=0.141(5) and 0.057(7) respectively.  For $S$=1 and 2 we would expect a total integral of 2.67 and 8 $\mu_{B}^{2}$ respectively.  Entropic arguments based upon high temperature heat capacity measurements would suggest that $S_{eff}$=${3 \over 2}$ is more appropriate and this would give a predicted integral of 5$\mu_{B}^{2}$, closer to our measurements given that even at 175 meV the top of the band has not been reached.   While our results are consistent with earlier low-energy measurements on Fe$_{1.11}$Te~\cite{Zaliznyak11:107}, we find significant spectral weight extending up to high energies giving the apparent low-temperature discrepancy.  More discussion on this point is provided in the supplementary information.  The integrated intensities are difficult to understand in terms of a purely localized model with $S$=1 or 2 discussed above, therefore suggesting the importance of itinerant effects.  Such effects maybe captured by considering orbital transitions~\cite{Yin13:xx} or an orbitally entangled ground state which can also account for the highly anisotropic nature suggested by the high energy spin dynamics.~\cite{Chaloupka13:110}  We have searched for high energy orbital transitions without success and this is discussed in the supplementary information (see also Ref. \onlinecite{Kim11:84,Cowley13:88,Hill08:100,Perkins98:58,Stock10:81}).

In summary, our work finds three results based upon our study the spin fluctuations in parent Fe$_{1+x}$Te.  First, we observe the presence of a soft incommensurate stripy excitations.  Second, by applying sum rules, we find the integrated intensity to be inconsistent with a $S$=1 ground state expected in the presence of a strong crystalline electric field.   Third, our results produce an apparent hour-glass structure which defines a cross over point from two dimensional fluctuations to one dimensional.  The results point to the parent Fe$^{2+}$ ground state of chalcogenide superconductors being highly anisotropic and also in an intermediate state between strong ($S$=2) and weak ($S$=1) orbital ground states.   

We are grateful for funding from the RSE, the Carnegie Trust, STFC, and through the NSF (DMR-09447720).  


%

\end{document}


\title{Supplementary information for ``Soft striped magnetic fluctuations competing with superconductivity in Fe$_{1+x}$Te"}

\author{C. Stock}
\affiliation{School of Physics and Astronomy, University of Edinburgh, Edinburgh EH9 3JZ, UK}
\author{E. E. Rodriguez}
\affiliation{Department of Chemistry of Biochemistry, University of Maryland, College Park, MD, 20742, U.S.A.}
\author{O. Sobolev}
\affiliation{Forschungs-Neutronenquelle Heinz Maier-Leibnitz, FRM2 Garching, 85747, Germany}
\author{J. A. Rodriguez-Rivera}
\affiliation{NIST Center for Neutron Research, National Institute of Standards and Technology, 100 Bureau Dr., Gaithersburg, MD 20889}
\affiliation{Department of Materials Science, University of Maryland, College Park, MD  20742}
\author{R.A. Ewings}
\affiliation{ISIS Facility, Rutherford Appleton Laboratory, Didcot, OX11 0QX, U.K.}
\author{J.W. Taylor}
\affiliation{ISIS Facility, Rutherford Appleton Laboratory, Didcot, OX11 0QX, U.K.}
\author{A. D. Christianson}
\affiliation{Quantum Condensed Matter Division, Oak Ridge National Laboratory, Oak Ridge, Tennessee 37831, USA}
\author{M. A. Green}
\affiliation{School of Physical Sciences, University of Kent, Canterbury, CT2 7NH, UK}

\date{\today}

\begin{abstract}

We present supplementary information regarding the experimental details, ``spurious" phonon scattering, and the formalism used to derive absolute units for the neutron scattering intensities.  In particular, we discuss the zeroeth moment sum rule presented in the main text of the paper and how we compared our data against this constraint on neutron scattering.  More discussion regarding the comparison with other neutron inelastic scattering experiments is also expanded.  We then discuss low-energy phonons which potentially overlap measurements of the magnetic excitations - particularly at low energies below $\sim$ 30 meV.  By combining spallation time-of-flight and reactor triple-axis we have checked for consistency and taken steps to ensure such phonon contamination is absent in the data presented in the main text.

\end{abstract}

\maketitle

\section{Sample details}

The electronic and magnetic properties in Fe$_{1+x}$Te$_{1-y}$Ch$_{y}$ can be tuned through either anion or cation substitution and has been the topic of several detailed studies.~\cite{Rodriguez10:132,Rodriguez11:2,Rodriguez11:84,Rodriguez13:88,Stock11:84,Stock12:85,Sales09:79,Rossler11:84,Koz13:88}  The results are generally consistent for the specific case of interstitial iron doping through tuning the variable $x$.  Two extremes of the phase diagram are reported with a critical concentration of $x\sim$ 10-12 \% separating commensurate collinear magnetic order at low concentrations from incommensurate spiral order at large interstitial iron concentrations.  We note that a previous diffraction study reported in Ref. \onlinecite{Zaliznyak12:85} was performed near this critical concentration and hence observed both commensurate and incommensurate orders.  Ref. \onlinecite{Mizuguchi12:152} reproduces the phase diagram found by ourselves~\cite{Rodriguez11:84,Rodriguez13:88} and in Refs. \onlinecite{Rossler11:84,Koz13:88} but with larger values of interstitial iron concentrations.  Diffraction studies discussing the critical properties near this critical concentration of interstitial iron have been reported in Ref. \onlinecite{Rodriguez11:84,Rodriguez13:88}. The current study reported in the main text discusses the spin fluctuations in both of these extremes and finds strong differences across the entire dispersion band.

The sample preparation and details for single crystals used in this study have been discussed in detail in several previous studies.~\cite{Rodriguez13:88,Rodriguez11:84}  Both $\sim$ 6 g samples were prepared by the Bridgeman technique and have been characterized by electrical resistivity, heat capacity, and neutron diffraction (polarized, unpolarized, and powder).  Low-energy neutron inelastic scattering data has also been reported for both samples.~\cite{Stock10:82}   

\section{Experimental details}

The magnetic response in Fe$_{1+x}$Te possesses a highly three-dimensional line shape in momentum at low energies which crosses over to a more two-dimensional line shape at higher energies.  To track the momentum dependence, we have used a combination of three-axis measurements performed at reactor based sources combined with spallation source data taken at higher energies.  While the momentum transfer of all measurements performed with a three-axis spectrometer was tuned to a particular value with all three components of the momentum transfer vector fixed, the spallation data was taken with the $c$-axis component (or the $L$ direction) being an implicit variable coupled, and hence varying, with energy transfer and in-plane momentum.  This technique works for purely two-dimensional or one-dimensional systems, but care needs to be taken for three dimensional excitations.  Three different sets of experiments were performed on each of the two interstitial iron concentrations $x$ discussed in the main text.  We outline the detailed experimental setups used here.  

\subsection{Cold triple-axis measurements - MACS}

Low energy (E$<$10 meV) measurements were performed using the MACS (Multi Axis Crystal Spectrometer located on the NG0 guide position) at the NIST Center for Neutron research.   Constant energy planes were constructed by fixing the final energy to E$_{f}$=3.6 meV using the 20 double-bounce PG(002) analyzing crystals and detectors and varying the incident energy by a double-focused PG(002) monochromator.  Each detector channel was collimated using 90$'$ Soller slits before the analyzing crystal.  Further details of the instrument design can be found in Ref. \onlinecite{Rodriguez08:19}. Full maps of the spin excitations in the (H0L) scattering plane, as a function of energy transfer, were then constructed by measuring a series of constant energy planes.  Warm Beryllium filters were used on the scattered side to filter out higher order neutrons.  All of the data has been corrected for the $\lambda/2$ contamination of the incident-beam monitor, and an empty cryostat background has been subtracted.  The sample was aligned in the (H0L) scattering plane for measurements performed on MACS.

\subsection{Thermal triple-axis measurements - HB1 (ORNL) and PUMA (FRM2)}

Medium energy transfers (E=5-20 meV) were performed using thermal triple-axis spectrometers at Oak Ridge National Labs (HB1) and the FRM2 reactor (PUMA).  On PUMA, a vertically focussing and horizontally flat PG(002) monochromator was used with a horizontally flat PG(002) analyzer.  Soller slit collimators were used and the sequence were fixed at 40$'$-mono-40$'$-$S$-open-analyzer-open.  The experiments used a fixed final energy of 13.5 meV and a PG filter was placed on the scattered side to remove higher-order contamination from the monochromator.  The choice was made to not use horizontal focussing on the monochromator and analyzer due to the presence of the close proximity of phonons which were found to easily mimic magnon dispersion and contaminate the results - particularly above $\sim$ 20 meV.  Further discussion is given below.  On HB1, a vertically focussed and horizontally flat PG(002) monochromator was used with a horizontally and vertically flat PG(002) analyzer.  The final energy was fixed at E$_{f}$=13.5 meV and a graphite filter was used on the scattered side to remove higher order neutrons from the monochromator.  Soller slit collimation of 80$'$ was used before and after the sample position.  All data from these thermal triple-axis experiments have been corrected for contamination of the incident beam monitor using calculations described in the appendix of Ref. \onlinecite{Stock04:69} and in Ref. \onlinecite{Shirane:book}.   For the PUMA and most of the HB1 measurements the sample was aligned in the (H0L) scattering plane and some measurements, described below, were done in the (HHL) scattering on HB1.  
 
\subsection{Spallation time-of-flight measurements - MAPS}

Higher energy transfers overlapping with the dynamic ranges discussed above were performed using the MAPS chopper spectrometer at the ISIS Facility.   The sample was aligned such that Bragg positions of the form (H0L) lay within the horizontal scattering plane and cooled with a bottom loading closed cycle refrigerator to temperatures of 10 K.  The $c$ axis was aligned parallel to the incident beam ($\vec{k}_{i}$).  The $t_{0}$ chopper was spun at a frequency of 50 Hz and phased to remove high energy neutrons from the target.  A ``sloppy" Fermi chopper was used to monochromate the incident beam.  To cover a wider dynamic range, we used three different incident energies of E$_{i}$=75, 150, and 300 meV with the Fermi chopper spun at frequencies of 200, 250, 350 Hz respectively.  The energy resolution at the elastic ($E$=0) position was 4.0, 9.0, and 18.1 meV for E$_{i}$=75, 150, and 300 meV configurations respectively.  The detectors consisted of an array of position sensitive detectors allowing good angular resolution both within and vertical to the scattering plane.

\subsection{Spallation time-of-flight measurements - MARI}

Searches for high-energy crystal field excitations, which may result from an orbital degree of freedom, were performed on the MARI direct geometry spectrometer.  The experiments used a t$_{0}$ chopper spun at 50 Hz to remove high energy neutrons in parallel with a ``relaxed" Fermi chopper spun at 600 Hz.  The sample was cooled in a closed cycle cryostat to temperatures between 3 K and 400 K and the incident beam was aligned parallel to the $c$ axis.  The configuration was used assuming single-ion type excitations which have no strong dispersion in momentum.  The results from this experiment are described below.

\section{Neutron cross sections, absolute normalization, and sum rules}

In this section we outline the cross sections and normalization methods used to obtain the data in the main text of the paper.  We note that similar analyses have been performed and discussed for the cuprates.~\cite{Fong00:61,Dai13:88}  While several papers outline the formalism and the methods for this we largely follow the formulae and methods used Ref. \onlinecite{Stock04:69}.

\subsection{Cross sections and absolute intensities}

For completeness, we outline the cross sections used in this section for obtaining the integrated intensities quoted in the main text.  We have performed this analysis so that our results can be compared with other systems and with theory.  We outline the formulae for the case of triple-axis as the spallation data taken on MAPS was directly normalized using a vanadium standard.  In the case of a triple-axis spectrometer measured with an incident beam monitor, the measured intensity from phonon or magnetic scattering is directly proportional to the magnetic or phonon cross section.  We note that a detailed discussion of this process is described in Ref. \onlinecite{Xu13:84}.

\begin{eqnarray}
I_{ph,mag}(\vec{Q},E) \propto S_{ph,mag}(\vec{Q},E).
\label{cross_section}
\end{eqnarray}

The constant of proportionality ($A$) can be determined from the measured integrated intensity of a phonon or using a known vanadium standard.  In the case of a phonon, the cross section is 

\begin{eqnarray}
I_{ph}(\vec{Q})=A \left( {\hbar \over {2\Omega_{0}}} \right) [1+n(E)] |F_{N}|^{2} ...\\
 \times {{Q^{2} \cos^{2}(\beta)} \over M } e^{-2W} \nonumber
\label{cross_section}
\end{eqnarray}

\noindent where $M$ is the mass of the unit cell, the Debye-Waller factor $e^{-2W} \sim 1$, $[1+n(E)]$ is the Bose factor, $|F_{N}|^{2}$ is the static structure factor of the nearby Bragg reflection, $\Omega_{0}$ is the phonon frequency, and $\beta$ is the angle between $\vec{Q}$ and the phonon eigenvector.  Measuring the energy integrated intensity $I_{ph}(\vec{Q})=\int dE I_{ph}(\vec{Q},E)$ of the acoustic phonon therefore affords a measurement of the calibration constant $A$.

For the magnetic scattering, the magnetic correlation function is related to magnetic $S_{mag}(\vec{Q},E)$ by the following, 

\begin{eqnarray}
S_{mag}(\vec{Q},E)=g^{2}f^{2}(\vec{Q}) ... \\
\times \sum_{\alpha \beta} \left( \delta_{\alpha \beta} - \hat{Q}_{\alpha}\hat{Q}_{\beta} \right)S_{\alpha\beta}(\vec{Q},E). \nonumber
\label{cross_section}
\end{eqnarray}

\noindent The correlation function is related by the fluctuation dissipation theorem to the imaginary part of the spin susceptibility $\chi(\vec{Q},E)$ by

\begin{eqnarray}
S_{\alpha\beta}(\vec{Q},E)=\pi^{-1}[n(E)+1] {\chi''(\vec{Q},E)\over {g^{2}\mu_{B}^{2}} }.
\label{cross_section}
\end{eqnarray}

\noindent In our analysis, we assumed that the paramagnetic scattering is isotropic in spin, therefore, $\chi''=\chi''_{xx}=\chi''_{yy}=\chi''_{zz}$.    Putting this all together, we can then write the following for the magnetic cross section.

\begin{eqnarray}
I_{mag}(\vec{Q},E) = A {{\left(\gamma r_{0} \right)^{2}} \over 4} f^{2}(Q)e^{-2W} {[1+n(E)] \over {\pi \mu_{B}^{2}}}... \\
\times \left( 2 \chi''\right), \nonumber
\label{cross_section}
\end{eqnarray}

\noindent where ${{\left(\gamma r_{0}\right)^{2}} \over 4}$ is 73 mbarns sr$^{-1}$ and $f(Q)$ is the isotropic magnetic form factor for Fe$^{2+}$.  We emphasize here that we have assumed isotropic or paramagnetic scattering to fix the form of the spin susceptibility $\chi''$.  While it is likely that this approximation holds for our high-energy data where the energy transfer is much larger than the anisotropy gap, this assumption could potentially introduce errors into the spectral weight at low energies near the $\sim$ 6-7 meV energy gap.  Given the large amount of spectral weight that resides at high-energies, any error in the total integrated intensity is likely small in comparison to other uncertainties introduced through absolute normalization.

For thermal and cold triple-axis measurements, we have used an acoustic phonon for the absolute calibration and for spallation higher energy data (above $\sim$ 20 meV) we have used a vanadium standard.  For this calibration we have taken a cut through the elastic line assuming a dominant incoherent cross section of the vanadium standard.   When compared with the phonon calibration described above at lower energy transfers, both methods agreed within error.

\subsection{Zeroeth moment sum rule}

The intensity integrated overall energies and momentum transfer is constrained by the zeroeth moment sum rule.  This rum rule depends on the underlying value of the spin magnitude $S$ and therefore provides a means of understanding the ground state properties.  Here we write the equations used in the analysis discussed in the main text.

Integrating $S(\vec{Q},E)$ over all energies and momentum transfers is constrained by the following equation,

\begin{eqnarray}
\int dE \int d^{3}Q S_{mag}(\vec{Q},E)={2\over 3} g^{2} S(S+1).
\label{sum_rule}
\end{eqnarray}

\noindent Substituting in the cross section for paramagnetic scattering discussed above, we get the following,

\begin{eqnarray}
\tilde{I}=\pi^{-1} \int dE \int d^{3}Q [n(E)+1]\chi''(Q,E)=...\\ \nonumber
{1\over 3} g^{2} \mu_{B}^{2} S(S+1)
\label{sum_rule}
\end{eqnarray}

\noindent The integral is overall energy transfers including elastic ($E=0$) and inelastic contributions.  In our experiments, we were able to obtain reliable data up to 175 meV and therefore we have cut the integral arbitrarily at this value.    A breakdown of the total integrated spectral weight listing our measured dynamic (inelastic), static (elastic), and total values is provided in Table \ref{table_absolute}.

\begin{table}[ht]
\caption{Absolute intensities}
\centering
\begin{tabular} {c c c c}
\hline\hline
x & $\tilde{I}_{dynamic}$  ($\mu_{B}^{2}$) & $\tilde{I}_{static}$ ($\mu_{B}$) &  $\tilde{I}_{total}$ ($\mu_{B}^{2}$) \\
\hline\hline
0.057(7) & 0.49  & 1.8  & 3.7  \\
0.141(5) & 0.83 & 1.6  & 3.4  \\
\hline
\label{table_absolute}
\end{tabular}
\end{table}

Based on this analysis, we observe a significant amount of the low-temperature spectral weight is present in the inelastic channel.  For comparison, a similar analysis over a similar energy range in the cuprates (with $S$=${1\over 2}$) gave a total integral of $\sim$ 0.3 $\mu_{B}^{2}$ in ortho-II YBa$_{2}$Cu$_{3}$O$_{6.5}$.~\cite{Stock05:71}  We obtain consist results with Ref. \onlinecite{Zaliznyak11:107} when only integrating the data up to $\sim$ 30 meV - the same energy range probed in that experiment.  However, we observe significant spectral weight at higher energies up to 175 meV, not investigated in previous works.  We therefore do not find the low temperature results can be interpreted in terms of a S=1 strong crystal field framework as discussed in the main text.

\section{Weak dispersion along the $c$-axis}

\renewcommand{\thefigure}{S1}
\begin{figure}
\includegraphics[width=9cm] {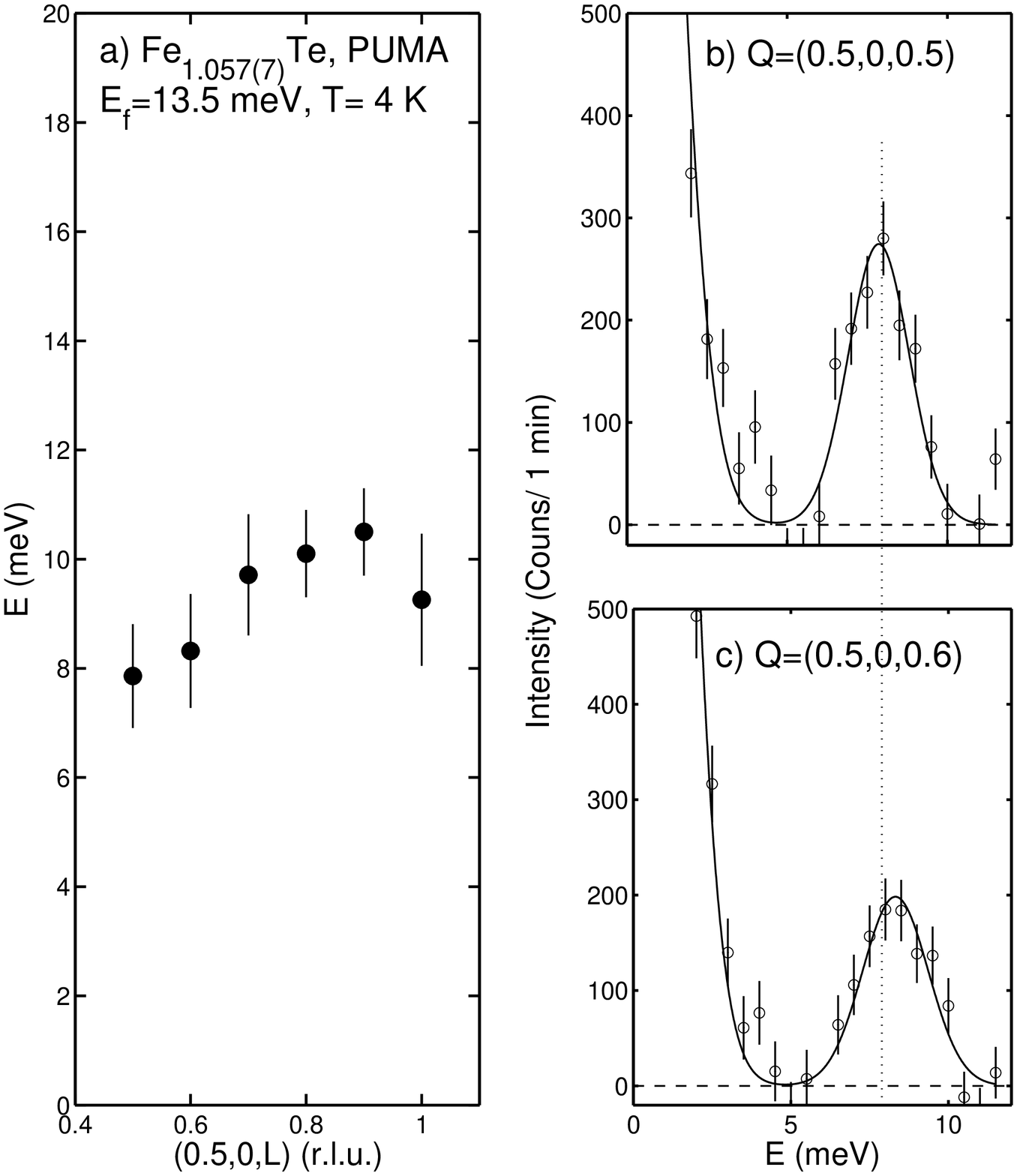}
\caption{\label{c_dispersion} $a)$ The $c$-axis dispersion taken from a series of constant-$Q$ scans examples of which are illustrated in $b)$ and $c)$.  The data illustrate a weak dispersion of the excitations along the $c$.}
\end{figure}

As noted above, the magnetic excitations at low-energies are three-dimensional in the sense that they are peaked in momentum along all three directions.   While the lower energy data was obtained using cold and thermal triple-axis spectrometers where the momentum transfer could be tuned to a particular position, the spallation source data depend on the excitations being two dimensional and therefore the value of $L$ on MAPS varies with energy transfer (see discussion in Ref. \onlinecite{Stock05:71}).  

To check this and over which energy range this approximation is valid, we have measured the $c$-axis dispersion in detail on the commensurate Fe$_{1.057(7)}$Te crystal using the PUMA thermal triple-axis spectrometer.  The results are illustrated in Fig. \ref{c_dispersion} taken at 4 K for momentum positions along $\vec{Q}$=(${1\over 2}$,0,0.5) to (${1\over 2}$,0,1.0).  Panel $a)$ shows the results of gaussian fits plotting the peak position as a function of momentum transfer.  The data show a very weak dispersion from $\sim$ 7 meV up to $\sim$ 10 meV consistent with results presented previously in Ref. \onlinecite{Stock11:84}.  Example scans are presented in panels $b)$ and $c)$. The vertical dashed line illustrates the comparatively small change in the frequency as the momentum transfer is changed along $L$.  Based on this result of a weak $c$-axis dispersion we consider the approximation of two-dimensional spin excitations to be valid over the energy range probed in our spallation source data.

\section{Stripe correlations and one dimensional scattering}

\renewcommand{\thefigure}{S2}
\begin{figure}
\includegraphics[width=8.5cm] {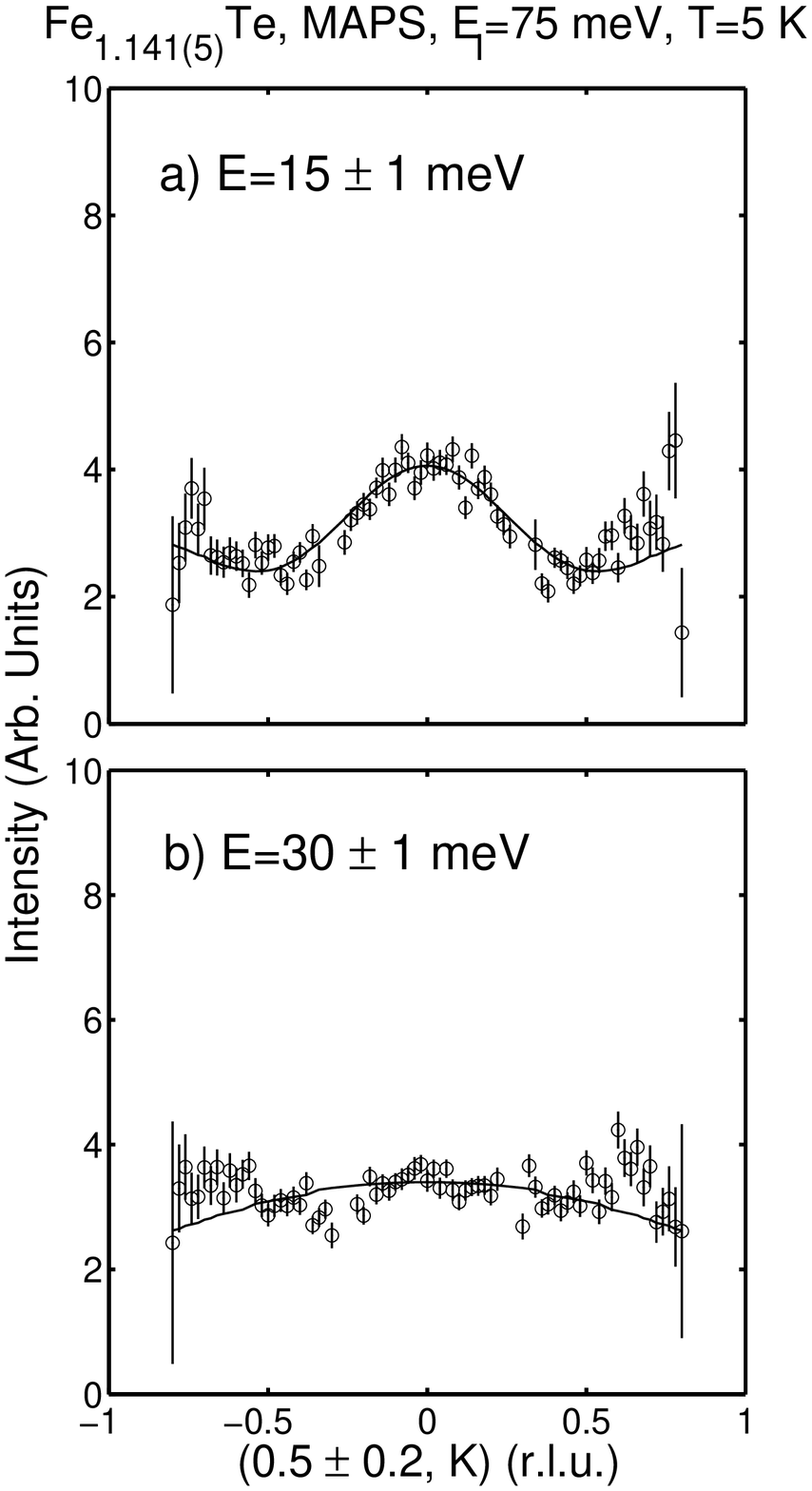}
\caption{\label{k_dispersion} The momentum dependence along K for Fe$_{1.141(5)}$Te at $a)$ 15 $\pm$ 1 meV and $b)$ 30 $\pm$ 1 meV.  The solid lines are a fit to the function described in the text.}
\end{figure}

As noted based on the two-dimensional slices presented in Fig. 1 of the main text, the inward dispersion or ``hourglass" dispersion marks the cross over from two dimensional excitations to one dimensional.  This was quantified as a function of energy for both interstitial iron concentrations by fitting the K dependence to the form $F(K)\propto  (1+2\alpha \cos(\vec{Q}\cdot \vec{b}))$ multiplied by the iron factor squared as discussed above to capture the fact that L is changing as a function of K.  Sample fits are shown in Fig. \ref{k_dispersion} illustrated weak correlations along K at low energies (panel $a$ where E=15 $\pm$ 1 meV) and the loss of these correlations at high energies (panel $b$ where E=30 $\pm$ 1 meV).  

\section{Phonon contamination of the magnetic signal}

In this section we discuss the possibility for phonon contamination of the magnetic signal in Fe$_{1+x}$Te and steps we have taken to avoid and check for this in our data.  

\subsection{Low-energy phonons and possible magnetic contamination giving $H$=0 scattering}

\renewcommand{\thefigure}{S3}
\begin{figure}
\includegraphics[width=9cm] {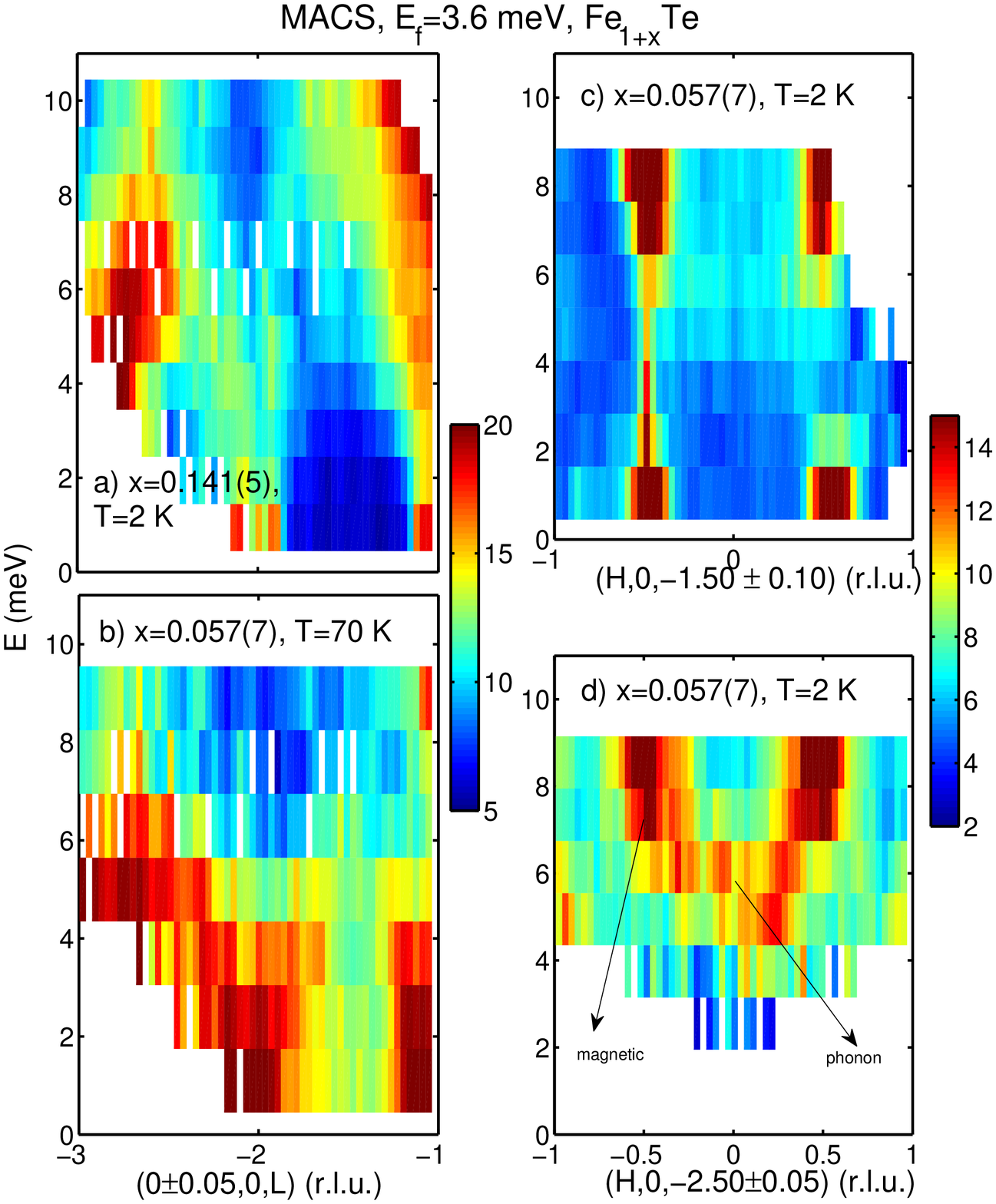}
\caption{\label{phonons_lowE} A summary of the low-energy $c$-axis propagating phonons based upon cold triple-axis data taken on MACS with E$_{f}$=3.6 meV.  $a)$ Illustrates a constant energy slice taken along the ($\pm$ 0.05, 0, L) direction at T=2 K for the incommensurate (spiral magnetic structure) Fe$_{1.141(5)}$Te crystal.  $b)$ Illustrates a similar slice taken for commensurate (collinear magnetic structure) Fe$_{1.057(7)}$Te.}
\end{figure}

During the course of the experiments reported in the main text, we discovered several spurious excitations near the dynamic magnetic scattering which later turned out to be due to phonons.  In this section, we report on several low-energy phonons up to 30 meV energy transfer that potentially contaminate the magnetic results.   We have checked that the scattering and in particular the dispersion relation in the main text is magnetic by comparing several different Brillouin zones.  Because of the low-energy phonons, we found spallation data was the most reliable between the energy ranges of 15- 30 meV and this was cross checked with thermal triple-axis measurements.  

Fe$_{1+x}$Te is a highly two dimensional material with layers weakly held together by van der Waals forces.  This is evidenced by the weak magnetic interactions along the $c$-axis discussed and characterized in the previous section.  One phonon branch which gave the appearance of commensurate (H=0) scattering in our experiments was a low-energy acoustic branch propagating along $c$.  This is illustrated in Fig. \ref{phonons_lowE} taken on MACS with E$_{f}$=3.6 meV.  Panel $a)$ illustrates a constant-$Q$ slice taken along the ($\pm$ 0.05, 0, L) direction for the incommensurate Fe$_{1.141(5)}$Te sample illustrating an acoustic phonon mode with a top of the band at L=1.5 and 0.5 of $\sim$ 7 meV.  A similar scan performed for  Fe$_{1.057(7)}$Te at 70 K (panel $b$) shows the same phonon mode with possibly a slightly lower maximum energy.  The top of the band in the acoustic phonon is very similar to the energy of the gap in the commensurate Fe$_{1.057(7)}$Te sample shown in Fig. 1 in the main text and appears at $L$ equal to half integer positions.

The problem in terms of measuring magnetic scattering is further illustrated in Fig. \ref{phonons_lowE} panels $c)$ and $d)$ which shows constant Q slices for L=-1.5 and -2.5.  The growth in intensity near H=0 with increasing $|Q|$ supports the fact that the commensurate H=0 scattering originates from phonons and not magnetism. 

\subsection{Phonons at 20-30 meV and possible magnetic contamination giving strong dispersion}

\renewcommand{\thefigure}{S4}
\begin{figure}
\includegraphics[width=8cm] {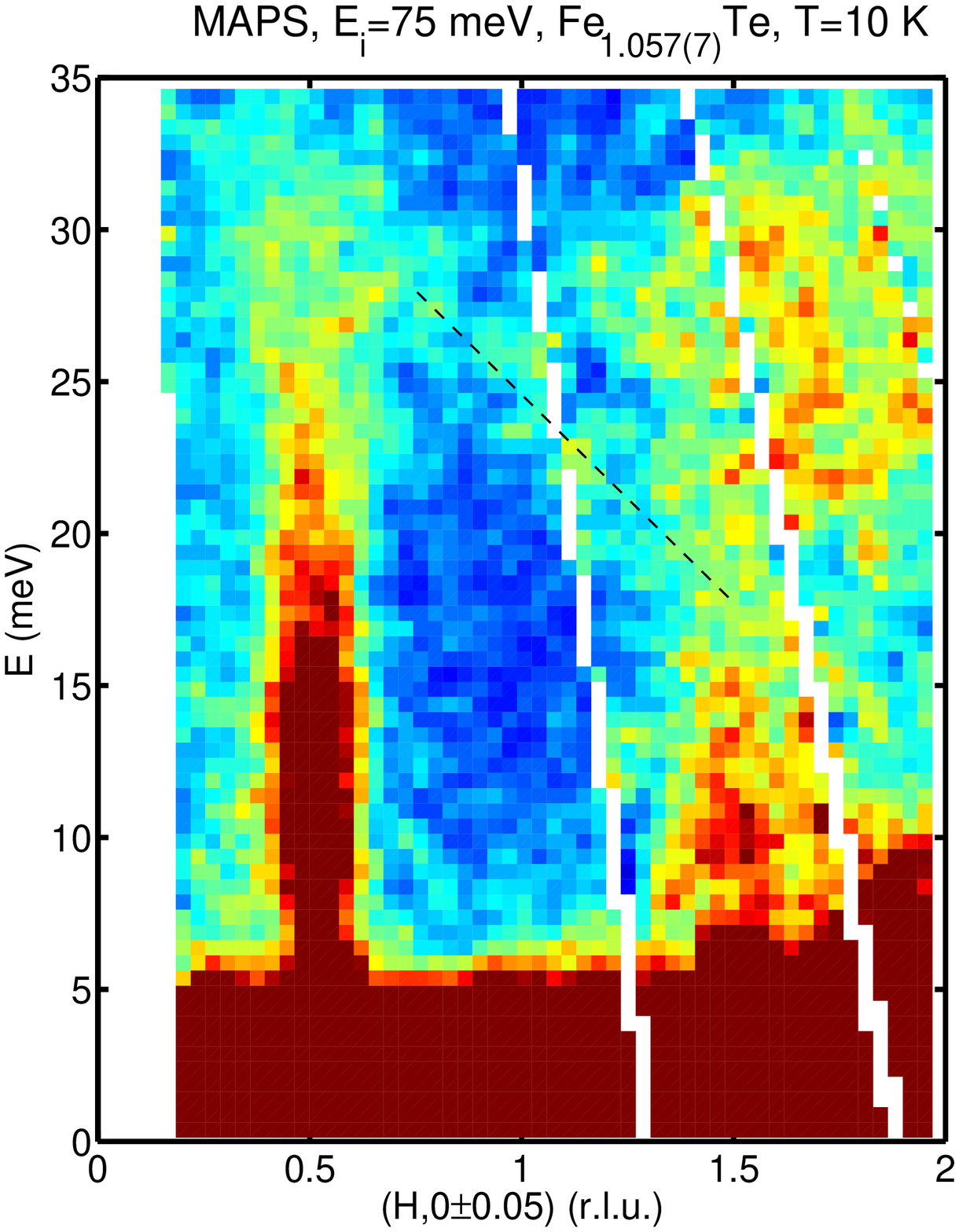}
\caption{\label{phonons_maps} A T=10 K constant-$Q$ slice illustrating a phonon branch highlighted by the dashed line.}
\end{figure}

The energy range extending from $\sim$ 20-30 meV was found to be difficult to study on thermal triple-axis spectrometers where large values of $L$ and $H$ were required to close the scattering triangle.  Similar to the situation above, we found there were several phonon branches which crossed the nuclear zone boundary giving the appearance of a strong dispersion in the magnetic scattering.  The problem is demonstrated in Fig. \ref{phonons_maps} where we show a constant-$Q$ slice from our MAPS data set on Fe$_{1.057(7)}$Te.  In this particular experiment, the $c$ axis was aligned along the incident beam k$_{i}$.  Note that while both in-plane momentum transfer components, $H$ and $K$, are well defined, the out of plane component $L$ varies with energy transfer.    An excitation branch crossing the magnetic scattering can be seen extending from $\sim$ 20-30 meV.

\renewcommand{\thefigure}{S5}
\begin{figure}
\includegraphics[width=8cm] {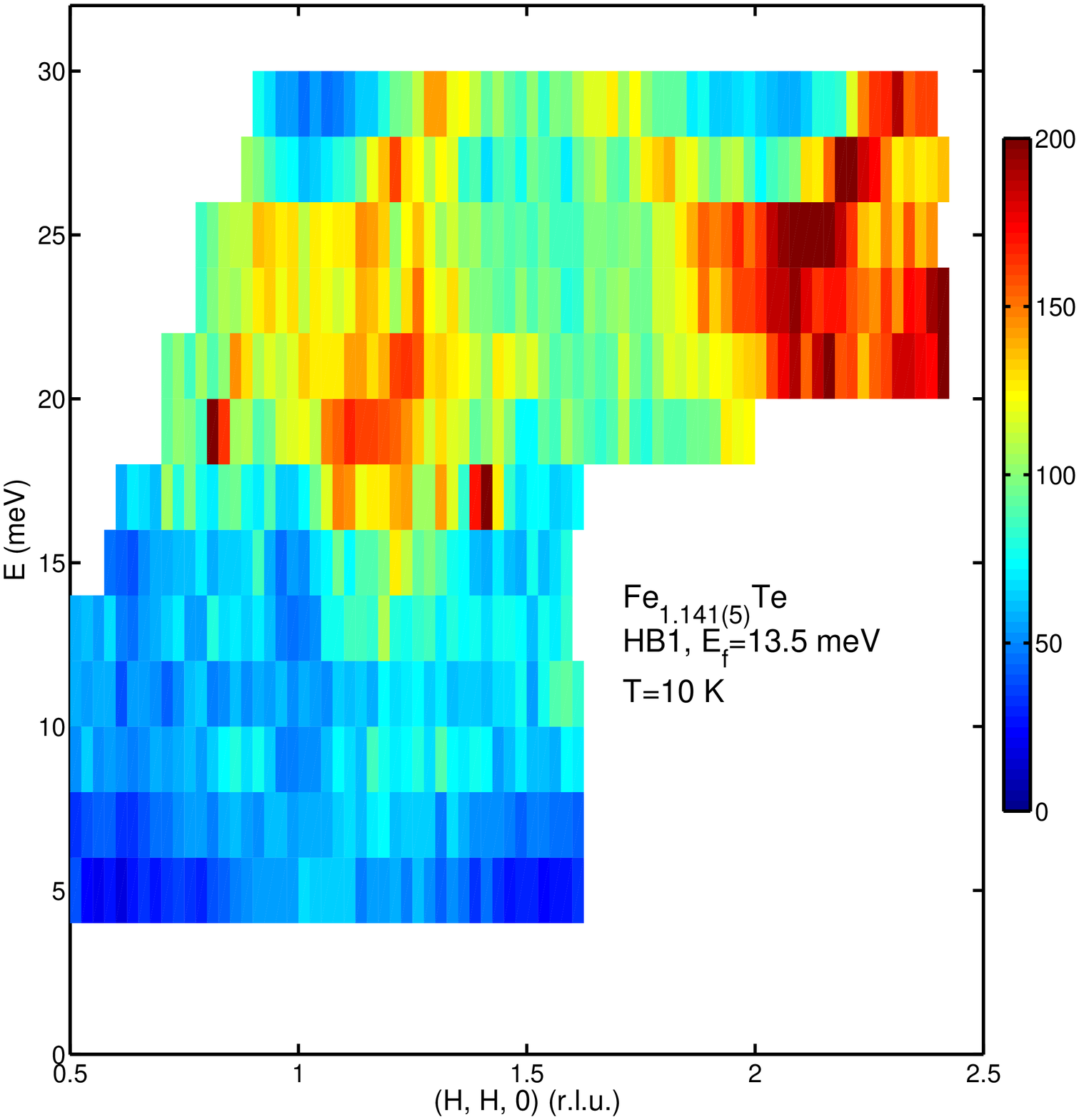}
\caption{\label{phonon_HB1} A constant-Q slice compiled from a series of constant energy scans on HB1 thermal triple-axis spectrometer (Oak Ridge).  A clear phonon branch extending in the range of 20-30 meV can be seen.}
\end{figure}

The phonon branch seen in Fig. \ref{phonons_maps} was also observed in our thermal triple-axis while tracking the magnetic dispersion in the range of 20-30 meV.  Figure \ref{phonon_HB1} shows a constant-Q slice compiled from a series of constant energy scans performed on Fe$_{1.141(5)}$Te in the (HHL) scattering plane.  The data show well defined phonons with a minimum of 20 meV and extending up to $\sim$ 30 meV where they cross the nuclear zone boundary but the magnetic zone centre.  These phonons contaminated several attempts both on PUMA and HB1 to extend the thermal triple-axis data into this range.    We therefore relied on spallation source data over this energy range to extract the magnetic dispersion reported in the main text.

\subsection{Hydrogen related modes contaminating results at high energies giving apparent crystal field levels}

Given the speculation and the large discussion around the issue whether Fe$^{2+}$ is in a S=1 or 2 ground state, we performed a search for higher energy orbital excitations in the range below $\sim$ 750 meV using the MARI direct geometry spectrometer at ISIS (configuration discussed above).   These searches were motivated by the possibility of spin-orbit transitions and the observation of similar orbital transitions using neutrons reported recently for Mott insulating NiO and CoO where a ground state orbital degeneracy exists.~\cite{Kim11:84,Cowley13:88}  There has also been the observation of similar high energy modes in superconducting YBa$_{2}$Cu$_{3}$O$_{6+x}$ which seemed to overlap with peaks observed using infrared and possibly inelastic resonant x-ray measurements.~\cite{Hill08:100,Perkins98:58}  The origin of these peaks in the cuprates is still unclear and as noted in Ref. \onlinecite{Stock10:82}, there are other scattering possibilities involving hydrogen related modes that occur over the same energy range.  Therefore checks need to be performed to determine the absence of hydrogen scattering and one test which can be performed on a wide-angle spectrometer like MARI is to search for a hydrogen recoil line as studied in detailed and demonstrated on polyethylene.~\cite{Stock10:81}  

\renewcommand{\thefigure}{S6}
\begin{figure}
\includegraphics[width=8cm] {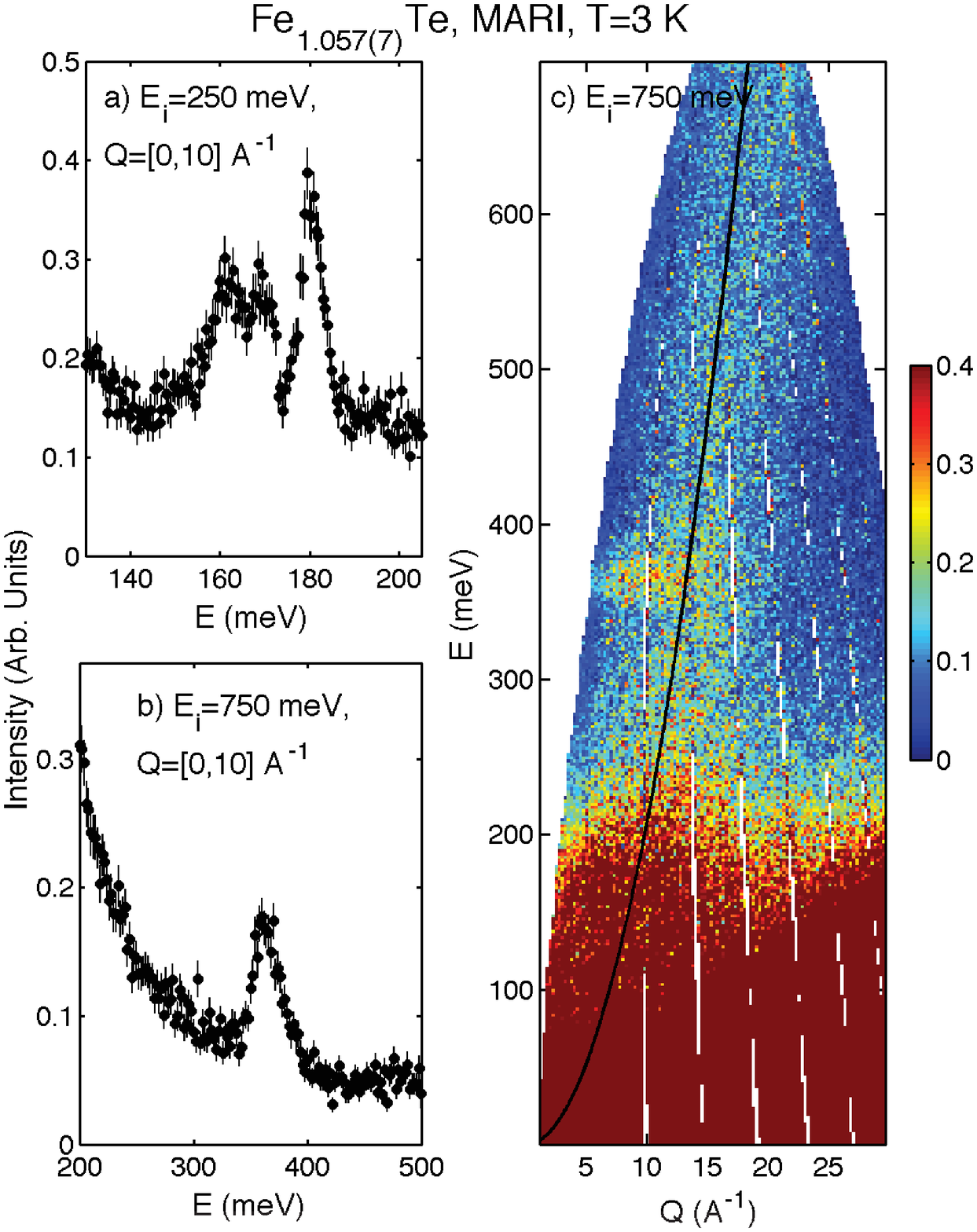}
\caption{\label{mari} High-incident energy scattering performed on MARI.  $a)$ and $b)$ illustrated one-dimensional cuts integrating over $Q$=[0,10] \AA$^{-1}$.  The results show several peaks at $\sim$ 170 meV and $\sim$ 370 meV.  $c)$ illustrates a constant-$Q$ slice taken with E$_{i}$=750 meV.  The solid curve shows the position of the expected hydrogen recoil line and the presence of intensity along this line indicates that the sample has absorbed hydrogen in some form.}
\end{figure}

We have performed such a test and the results are shown in Fig. \ref{mari}.  Panels $a)$ and $b)$ show that peaks are observed at low momentum transfer.  Panel $c)$ shows a representative constant-$Q$ slice showing the presence of an excitation at $\sim$ 370 meV.   The solid black line shows the predicted position of the hydrogen recoil line ($E_{recoil}={\hbar^{2} \over (2M_{p})}Q^{2}$ assuming the impulse approximation).  There are two problems with interpreting this excitation, and indeed the ones at $\sim$ 170 meV, as magnetic crystal field or orbital excitations.  First, the intensity of the excitations initially increases with momentum transfer and peaks at around the recoil line in momentum.  Second, as evidenced by intensity around the expected recoil position there is absorption of hydrogen into the sample.  Based on these two points we conclude that the sharp excitations represented in Fig. \ref{mari} $a)$ and $b)$ are due to hydrogen modes and not due to orbital transitions.   The width of these excitations was found not to respond to the structural and magnetic transition temperatures in this compound, but was observed to broaden at high temperatures near room temperature.  

Therefore, in summary, we conclude that the high-energy excitations observed are spurious and due to hydrogen absorbed into the sample.  The exact chemical origin of this remains unclear and we note that heating the sample at 400 K, while pumping, did not decrease the hydrogen recoil scattering implying that it originates beyond the surface of the sample and is not due to simply water absorption. 

\subsection{Conclusion from ``spurious" phonon scattering}

The first conclusion we draw from this analysis is that there is no measurable low-energy $H$=0 magnetic scattering in Fe$_{1+x}$Te. There is a low-energy $c$-axis propagating phonon mode which has a maximum energy position similar to the energy scale of the low-energy magnetic fluctuations.   The $magnetic$ scattering is therefore confined near the ($\pi$,0) position except at high energies where it disperse to the zone boundary as discussed in the main part of the text.  This phonon contamination needs to be accounted for and removed when considering the total moment and in particular the temperature dependence of the integrated intensity.  

The second conclusion is that the phonon scattering in the range near 20-30 meV crosses the magnetic zone centre and mimics a dispersion over this energy range.  We have avoided this problem by cross checking several different Brillouin zones and performing scattering experiments at the lowest Brillouin zones possible using chopper spectrometers at spallation sources.  We note that a similar problem was reported in the cuprates and was raised as a particular concern for triple-axis measurements.~\cite{Fong96:54}  In those cases detailed calculations were employed to check the data.  The development of new chopper spectrometers which can be used with high incident energies and small scattering angles allows a direct measurement of this.

The third conclusion is that the sharp excitations at high energies are the result of modes involving hydrogen scattering.  Again, the use of wide scattering angle chopper instruments with high incident energies allows a direct test for hydrogen in the sample.  


%